\documentclass[10pt,journal,twoside,twocolumn,nofonttune]{IEEEtran}

\usepackage{tabularx}
\usepackage{graphicx}
\usepackage{amsmath}
\usepackage{mathrsfs}
\usepackage{amsbsy}
\usepackage{amssymb}
\usepackage{mathtools}
\usepackage{upgreek}
\usepackage[twoside]{rotating}
\usepackage{psfrag}
\usepackage[numbers]{natbib}
\usepackage{color,soul}

\definecolor{LightGray}{gray}{0.8}
\definecolor{Orange}{rgb}{1.0, 0.31, 0.0}
\definecolor{Green}{rgb}{0.3, 1.0, 0.3}
\definecolor{Blue}{rgb}{0.75,0.75,1}

\newcommand{\fig}[1]{Fig.~\ref{#1}}

\newcommand{\tab}[1]{Table~\ref{#1}}
\newcommand{\tabs}[1]{Tables~\ref{#1}}

\newcommand{\bea}{\begin{eqnarray}}
\newcommand{\beal}[1]{\begin{eqnarray}\label{#1}}
\newcommand{\eea}{\end{eqnarray}}
\def\balg#1#2\ealg{\begin{align}\label{#1}#2\end{align}}
\def\balgnl#1\ealgnl{\begin{align*}#1\end{align*}}

\newcommand{\x}{{\mathbf r}}
\newcommand{\xt}{{\boldsymbol \rho}}

\renewcommand{\H}{{\mathbf H}}

\newcommand{\I}{{\mathbb I}}
\newcommand{\G}{{\mathbb G}}

\newcommand{\e}{\boldsymbol{\epsilon}}
\newcommand{\m}{\boldsymbol{\mu}}

\newcommand{\uphi}{{\mathbf e}_\varphi}

\newcommand{\uz}{{\mathbf e}_z}
\newcommand{\urho}{{\mathbf e}_\rho}
\newcommand{\un}{{\mathbf e}_n}

\begin{document}

\title{Modal Analysis of Gyrotropic Waveguides}

\author{Konstantinos Delimaris, \textit{Graduate Student Member, IEEE}, Georgios D. Kolezas, Carsten Rockstuhl,\\and Grigorios P. Zouros, \textit{Senior Member, IEEE}
\thanks{
Konstantinos Delimaris is with the School of Electrical and Computer Engineering, National Technical University of Athens, 15773 Athens, Greece; also with the Photonics Initiative, Advanced Science Research Center, City University of New York, New York, NY 10031, USA (e-mail: kdelimaris@gc.cuny.edu).

Georgios D. Kolezas and Grigorios P. Zouros are with the School of Electrical and Computer Engineering, National Technical University of Athens, 15773 Athens, Greece (e-mail: geokolezas@central.ntua.gr; zouros@ieee.org).

Carsten Rockstuhl is with the Institute of Theoretical Solid State Physics, Karlsruhe Institute of Technology, 76131 Karlsruhe, Germany; also with the Institute of Nanotechnology, Karlsruhe Institute of Technology, 76131 Karlsruhe, Germany (e-mail: carsten.rockstuhl@kit.edu).
}}

\markboth{}%
{DELIMARIS \lowercase{{\itshape et al}}.: Modal Analysis of Gyrotropic Waveguides}

\maketitle

\begin{abstract}
We report on the modal analysis of open gyrotropic waveguides (GWs). The GWs consist of a non-circular gyrotropic (i.e., gyroelectric and gyromagnetic) core and an infinitely extending isotropic cladding. To solve this problem, we develop two independent full-wave methods. The first technique is an extended integral equation (EIE) method, an extension of a previously developed EIE used to calculate the propagation constants in composite dielectric-isotropic waveguides. The second technique is a Chebyshev expansion method (CEM). In both implementations, the electric and magnetic fields in the gyrotropic core are expanded in superpotential-based cylindrical vector wave functions (SUPER-CVWFs), recently developed for the problem of oblique multiple scattering by gyrotropic cylinders. Both techniques allow us to calculate the propagation constants in the general case, without approximations. Various non-circular gyrotropic waveguides are considered. The EIE and CEM results are validated against a commercial finite element solver, and the accuracy and computational performance of both methods are benchmarked. A microwave application is presented where the complex propagation constants of ferrite rods are computed in the presence of an external magnetic flux density bias. Our work advances the theory of propagation in open waveguides, from dielectric/isotropic to gyrotropic ones, and enables the design of contemporary waveguiding structures.
\end{abstract}


\IEEEpeerreviewmaketitle

\section{Introduction}

\IEEEPARstart{O}{pen} waveguides are essential components of modern microwave and photonic systems, with applications that span a wide range of fields, from monopulse radar \cite{Geiger2020-ks}, beam-driven radiation sources \cite{Galyamin2021-en}, high-speed transmission in terahertz networks \cite{Kawanishi2019-vv}, fiber-optic sensors for environmental, industrial and biomedical usage \cite{Elsherif2022-zr}, signal distortions at mm-wave frequencies \cite{Meyer2022-nl}, to microwave-photonic interferometers based on optical carrier-based microwave interferometry \cite{zhu_tia_ma_hua_24}, and low latency transmission systems \cite{Sagae2024-gj}.

Traditionally, cylindrical dielectric waveguides (DWs) are made from isotropic materials, such as silica. Depending on the application, the dielectric permittivity of the core may exhibit step-index \cite{Karchevskii2014-gx}, inverse-parabolic graded-index \cite{Ung2014-tq}, helically cladded step-index \cite{Singh1995-ww}, or a radially varying profile \cite{Lahart1999-nj}. Modern guiding configurations, however, exploit anisotropic materials so as to achieve exotic behavior, such as metamaterial optical fibers for below cut-off propagation \cite{Pratap2015-jk}, twisted anisotropic permittivity fibers supporting orbital angular momentum \cite{Barshak2015-nu} and optical vortices \cite{Barshak2021-bq} for data transmission, and highly birefringent doped silica glass structures for C-band propagation \cite{Michalik2021-ua}.

Rigorous modal analysis of non-circular isotropic DWs, has been extensively reported in the past via various methods. Indicative techniques include the elliptical wavefunctions for elliptically shaped rods \cite{Yeh1962-nk}, \cite{Zouros2012-hk}, integral representations of the fields for arbitrary cross-sectional cores \cite{Eyges1979-ah}, and the method of auxiliary sources for non-circular DWs \cite{Kouroublakis2022-jw}. More recently, composite non-circular dielectric structures have been examined via a null field method-surface equivalence principle approach \cite{Kouroublakis2025-cb}, and via a vectorial extended integral equation method \cite{Delimaris2025-pr}.

Although several approaches have been reported for the modal analysis of isotropic DWs, as outlined above, reports on the rigorous modal analysis of open, non-circular anisotropic waveguides are scarce. In the present context, a gyrotropic waveguide (GW) is an open waveguide with a gyrotropic core. Specifically, a gyroelectric waveguide (GEW) features a gyroelectric core, and a gyromagnetic waveguide (GMW) features a gyromagnetic core.

A purely theoretical study of circular GMWs, without numerical results, has been initially reported in \cite{Epstein1956-mr} using the superposition principle. The analysis of circular GEWs has been examined in \cite{Uzunoglu1981-yq} via a cylindrical eigenfunction expansion method in which, although the characteristic equation for the general case (i.e., without restriction on the tensorial permittivity elements of the gyroelectric core) is given, numerical results are limited to the case where the non-diagonal element is small, compared to the other elements. A modal theory on transversely open anisotropic waveguides has been presented in \cite{Lindell1983-wf} via a variational technique, open anisotropic waveguides with diagonal elements in the permittivity have been examined in \cite{Shi1989-kd} by a perturbation method, while permittivity perturbed open anisotropic waveguides have been analyzed in \cite{Rongqing1989-nd} by an equivalent current theory. The Mathieu functions method has been applied in \cite{kang} for the modal analysis of anisotropic elliptical open waveguides. The cylindrical eigenfunctions method has been also applied in \cite{Alexeyev2013-yw} for the analysis of optical vortices in open anisotropic waveguides. Circular and non-circular open anisotropic waveguides with multilayered uniaxial anisotropy have been studied in \cite{Kamra2019-si} by representing each layer via a hybrid matrix and then extracting an equivalent circuit of the structure. Finally, a transformation optics technique has been applied in \cite{Napiorkowski2021-bj} to model a twisted permittivity in open anisotropic waveguides. 

In this work, we report, for the first time, on the rigorous and full-wave modal analysis of both circular and non-circular GWs. We calculate the propagation constants in such waveguides by utilizing two independent full-wave methods. The first technique is based on the extended integral equation (EIE), an approach previously used to calculate the propagation constants in composite isotropic DWs \cite{Delimaris2025-pr}. The second method is a Chebyshev expansion method (CEM) that does not involve integrations, developed here by using Chebyshev polynomials to satisfy the boundary conditions (BCs) at the core-cladding interface. In both methods, the EIE and CEM, the electromagnetic (EM) field is expanded in the recently developed superpotential-based cylindrical vector wave functions (SUPER-CVWFs) \cite{Zouros2025-ok}. Such an expansion allows us to readily express the fields in a gyrotropic medium when an $\exp(i\beta z)$ dependence on $z$ holds, i.e., in cases in which $\partial/\partial z\rightarrow i\beta\neq 0$, where $\beta$ is the propagation constant. Each employed method yields, once the BCs are enforced, a determinantal equation, the roots of which are the propagation constants $\beta$.

The EIE and CEM are exhaustively validated by comparison with alternative techniques in a plethora of cases: (i) with the well-known separation of variables method (SVM) \cite{marcuse_2ed} for circular isotropic DWs; (ii) with the perturbation solution for circular GEWs \cite{Uzunoglu1981-yq}; (iii) with the full-wave commercial finite-element solver COMSOL for circular GEWs and GMWs in the general case, i.e, without restriction on the tensor elements of the core permittivity/permeability; and (iv) with COMSOL for non-circular GEWs and GMWs, including cores of elliptical, rounded-triangular, and rounded-square cross-section. The pros and cons of each method, along with their efficiency relative to COMSOL, are also discussed. Finally, to demonstrate the applicability of the EIE and CEM, a microwave application is studied in which the complex propagation constants of ferrite GMWs are computed in the presence of an external magnetic flux density bias.

The novelty of this work lies in the following points. (i) A complete, rigorous, full-wave modal analysis is given for both circular and non-circular GWs with gyroelectric and gyromagnetic cores. The present analysis constitutes a step forward in the study of open waveguides, from traditional isotropic cores to gyrotropic cores. In particular, regarding the circular GMW, the present work advances \cite{Epstein1956-mr} since only a theoretical discussion was given in \cite{Epstein1956-mr} without proper numerical results. In contrast, the current study includes a comprehensive discussion on the theory, analysis, validation, and an investigation of a microwave application regarding ferrite GMWs. Regarding the circular GEW, the present work advances \cite{Uzunoglu1981-yq} where a characteristic equation for the general case (i.e., without restriction on the tensorial permittivity elements of the gyroelectric core) is given, but numerical results are limited to the case where the non-diagonal element is small, as compared to the other elements. The current study has no such limitations. (ii) The acquired solution is full-wave. In addition, no restrictions apply to the values of the tensorial permittivity/permeability elements. For example, the solution is not limited to the weakly guiding approximation (WGA), in which the refractive index of the core is assumed to be similar to that of the cladding. (iii) Construction of two independent methods, i.e., the EIE and the CEM, for the calculation of the propagation constants $\beta$ when the gyrotropic core has a non-circular cross-section. (iv) The starting point in both EIE and CEM is based on the expansion of the fields in SUPER-CVWFs. Although the SUPER-CVWFs were developed in \cite{Zouros2025-ok} for a multiple-scattering problem, the present work uses these recently developed expressions to solve a different problem, namely that of hybrid wave propagation in GWs. Concerning the EIE, the present approach constitutes an extension of the EIE employed in \cite{Delimaris2025-pr}, where now the SUPER-CVWFs are used in the kernels of the integral representations (IRs), instead of the standard CVWFs that are extensively utilized for isotropic structures. New integral terms are now obtained and used to construct the determinantal equation to calculate the $\beta$ in GWs. On the other hand, the CEM is not based on integrations like the EIE. Instead, the BCs on the core-cladding interface are expanded in terms of Chebyshev polynomials. This improves calculation accuracy for non-circular cross-sections, where the EIE performs poorly.

\begin{figure}[!t]
	\centering
	\includegraphics[scale=1.0]{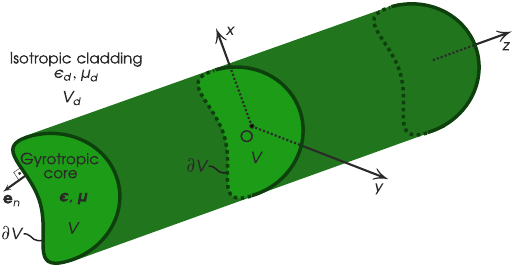}
	\caption{Configuration of the GW.}
	\label{geometry}
\end{figure}

The paper is organized as follows: Section~\ref{SOL} develops the solution, Section~\ref{VAL} discusses the validation, convergence, and performance of the methods, Section~\ref{APP} discusses a microwave application on ferrite GMWs, and Section~\ref{CON} concludes the paper. Finally, Appendices~\ref{APP_A} and \ref{APP_B} contain the integrals and the summations needed for the application of the EIE and the CEM, respectively.

\section{Solution of the Problem}\label{SOL}

\subsection{Configuration of the GW and the SUPER-CVWFs}

Figure~\ref{geometry} depicts the configuration of the non-circular GW. The gyrotropic core is defined by the boundary $\partial V$, with ${\mathbf e}_n$ being its outward-directed normal unit vector and $V$ its interior. Core $V$ is a medium with tensorial permittivity $\boldsymbol{\epsilon}$ and permeability $\boldsymbol{\mu}$, of the gyrotropic form
\balg{1}
\boldsymbol{\epsilon}=\epsilon_0\begin{bmatrix}
\epsilon &-i\epsilon_a&0\\
i\epsilon_a&\epsilon&0\\
0&0&\epsilon_z
\end{bmatrix},\,\,
\boldsymbol{\mu}=\mu_0\begin{bmatrix}
\mu &-i\mu_a&0\\
i\mu_a&\mu&0\\
0&0&\mu_z
\end{bmatrix}.
\ealg
In \eqref{1}, $\epsilon_0$ and $\mu_0$ are the permittivity and permeability of free space, while $\epsilon$, $\epsilon_a$, $\epsilon_z$, $\mu$, $\mu_a$, and $\mu_z$ are relative elements. The cladding in \fig{geometry}, denoted by $V_d$, is an isotropic dielectric with constitutive parameters $\epsilon_d=\epsilon_{dr}\epsilon_0$ and $\mu_d=\mu_{dr}\mu_0$.

For a propagation inside a gyrotropic medium, where $\partial/\partial z\neq 0$, the longitudinal components $E_z$ and $H_z$ of the electric and magnetic field are coupled \cite{Przeziecki1979-sm}, i.e, they do not separately satisfy homogeneous Helmholtz equations, as is the cutoff case when $\partial/\partial z= 0$. However, with the aid of the so-called superpotentials \cite{Przeziecki1979-sm}, analytical expressions can be found for the EM field components when $\partial/\partial z\rightarrow i\beta\neq 0$. In \cite{Zouros2025-ok}, we have developed the SUPER-CVWFs that enable an elegant vectorial expansion of the EM field inside a gyrotropic medium. These SUPER-CVWFs combine the standard CVWFs with the analytical expressions of the fields' components, as obtained from the theory of superpotentials. In particular, assuming an $\exp(-i\omega t)$ time dependence and $\partial/\partial z\rightarrow i\beta$, the electric field ${\mathbf E}(\x)$ inside $V$ can be expanded as
\balg{2}
{\mathbf E}(\x)=\sum_{j=1}^2\sum_{m=-\infty}^\infty A_{jm}\big[&\tilde{{\mathbf M}}_m^{(1)}(\chi_j,\x)\notag\\
&+\tilde{{\mathbf N}}_m^{(1)}(\chi_j,\x)+\tilde{{\mathbf L}}_m^{(1)}(\chi_j,\x)\big],
\ealg
where $\x=(\rho,\varphi,z)$ is the position vector in cylindrical coordinates, $A_{jm}$ are unknown expansion coefficients, and $\tilde{{\mathbf M}}_m^{(1)}(\chi_j,\x)$, $\tilde{{\mathbf N}}_m^{(1)}(\chi_j,\x)$, and $\tilde{{\mathbf L}}_m^{(1)}(\chi_j,\x)$ are the electric SUPER-CVWFs, having the expressions
\balg{3}
\tilde{{\mathbf M}}_m^{(1)}(\chi_j,\x)&=e^{i\beta z}e^{im\varphi} T_j\!\Big[i\frac{m}{\rho} J_{m}(\chi_j\rho)\urho\!-\!\chi_jJ'_m(\chi_j\rho)\uphi\Big] ,\notag\\
\tilde{{\mathbf N}}_m^{(1)}(\chi_j,\x)&=e^{i\beta z}e^{im\varphi}\frac{W_j-i\beta S_j}{\chi_j^2+\beta^2} \!\Big[i\beta\chi_j J'_m(\chi_j\rho)\urho\notag\\
&\quad\,-\!\frac{m\beta}{\rho}J_m(\chi_j\rho)\uphi\notag\!+\!\chi_j^2J_m(\chi_j\rho)\uz\Big],\notag\\
\tilde{{\mathbf L}}_m^{(1)}(\chi_j,\x)&=e^{i\beta z}e^{im\varphi}\frac{\chi_j^2S_j-i\beta W_j}{\chi_j^2+\beta^2} \!\Big[\chi_j J'_m(\chi_j\rho)\urho\notag\\
&\quad\,+\!i\frac{m}{\rho}J_m(\chi_j\rho)\uphi\!+\!i\beta J_m(\chi_j\rho)\uz\Big].
\ealg
In \eqref{3}, $\chi_j$ are given by \cite[eq.~(6)]{Zouros2025-ok}, $T_j$, $W_j$, and $S_j$ are given by \cite[eq.~(8)]{Zouros2025-ok}, $J_m(x)$ is the Bessel function and $J'_m(x)$ its derivative with respect to the argument. The magnetic field ${\mathbf H}(\x)$ inside $V$ is expanded by
\balg{4}
{\mathbf H}(\x)=\sum_{j=1}^2\sum_{m=-\infty}^\infty A_{jm}\big[&\hat{{\mathbf M}}_m^{(1)}(\chi_j,\x)\notag\\
&+\hat{{\mathbf N}}_m^{(1)}(\chi_j,\x)+\hat{{\mathbf L}}_m^{(1)}(\chi_j,\x)\big],
\ealg
where $\hat{{\mathbf M}}_m^{(1)}(\chi_j,\x)$, $\hat{{\mathbf N}}_m^{(1)}(\chi_j,\x)$, and $\hat{{\mathbf L}}_m^{(1)}(\chi_j,\x)$ are the magnetic SUPER-CVWFs, given by \eqref{3} by replacing $T_j\rightarrow N$, $W_j\rightarrow R_j$, and $S_j\rightarrow M_j$, with $N$, $R_j$, and $M_j$ given by \cite[eq.~(8)]{Zouros2025-ok}.

Inside the isotropic cladding $V_d$, the electric field ${\mathbf E}^d(\x)$ is expanded using the standard CVWFs, as
\balg{5}
{\mathbf E}^d(\x)=\sum_{m=-\infty}^\infty \big[B_{m}{\mathbf M}_m^{(6)}(\kappa_c,\x)+\Gamma_m{\mathbf N}_m^{(6)}(\kappa_c,\x)\big].
\ealg
In \eqref{5}, $B_m$ and $\Gamma_m$ are unknown expansion coefficients, $\kappa_c=(\beta^2-k_d^2)^{1/2}$, while $k_d=\omega(\epsilon_d\mu_d)^{1/2}$ is the wavenumber in $V_d$. The CVWF ${\mathbf M}_m^{(6)}(\kappa_c,\x)$ is given by $T_j^{-1}\tilde{{\mathbf M}}_m^{(1)}(\chi_j,\x)$ with $\chi_j\rightarrow\kappa_c$, $J_m(x)\rightarrow K_m(x)$ the modified Bessel function of the second kind that represents decaying waves when $\kappa_c$ real, and $J'_m(x)\rightarrow K'_m(x)$, i.e., the derivative of $K_m(x)$ with respect to the argument. The CVWF ${\mathbf N}_m^{(6)}(\kappa_c,\x)$ is given by $k_d^{-1}(\chi_j^2+\beta^2)/(W_j-i\beta S_j)\tilde{{\mathbf N}}_m^{(1)}(\chi_j,\x)$ with $\chi_j\rightarrow\kappa_c$, $\chi_j^2\rightarrow-\kappa_c^2$, $J_m(x)\rightarrow K_m(x)$, and $J'_m(x)\rightarrow K'_m(x)$. The magnetic field ${\mathbf H}^d(\x)$ inside $V_d$ is obtained by ${\mathbf H}^d(\x)=-i/(\omega\mu_d)\nabla\times{\mathbf E}^d(\x)$.

In the remainder of this Section, we report on the full-wave modal analysis of circular GWs by applying the SVM, and of non-circular GWs by utilizing the EIE and the CEM. In what follows, the EM field is expressed by ${\mathbf F}(\x)={\mathbf F}(\xt)e^{i\beta z}$, ${\mathbf F}={\mathbf E},{\mathbf H}$, where $\xt=(\rho,\varphi)$ is the position vector in polar coordinates. The non-circular core is defined by assuming, on the transverse $xy$-plane, the polar coordinate system $(\rho,\varphi)$ with origin ${\rm O}$, as depicted in \fig{geometry}. Then, $\partial V$ is defined by the polar equation $\rho=\rho(\varphi)$, $0\leqslant\varphi<2\pi$. For a circular core, $\rho=a$, with $a$ being the radius.

\subsection{Circular GW via the SVM}\label{SOL-B}

For the modal analysis of circular cores, we apply the SVM. To construct the determinantal equation, we satisfy the BCs $\urho\times[{\mathbf E}^d(a,\varphi,z)-{\mathbf E}(a,\varphi,z)]=0$ and $\urho\times[{\mathbf H}^d(a,\varphi,z)-{\mathbf H}(a,\varphi,z)]=0$. Substituting \eqref{2}, \eqref{4}, \eqref{5}, and the expansion for ${\mathbf H}^d(\xt)=-i/(\omega\mu_d)\nabla\times{\mathbf E}^d(\xt)$ using \eqref{5}, and utilizing the orthogonality relations of the exponential functions, we obtain, exactly, the following homogeneous system for the full-wave calculation of the $\beta$, i.e., $\mathbb{Q}(\beta){\mathbf u}=0$, where the vector ${\mathbf u}=[A_{1m}, A_{2m}, B_m, \Gamma_m]^T$ and the matrix $\mathbb{Q}(\beta)=(q^{pq}_m(\beta))\in\mathbb{C}^{4\times4}$, $p,q=1,2,3,4$, with elements $q^{1j}_m=-T_j\chi_jJ'_m\left(\chi_j a\right) + \left(i m S_j/a\right)J_m\left(\chi_j a\right)$, $j=1,2$, $q^{13}_m = \kappa_c K'_m\left(\kappa_c a\right)$, $q^{14}_m = [m\beta/\left(k_da\right)]K_m\left(\kappa_c a\right)$, $q^{2j}_m=W_jJ_m\left(\chi_j a\right)$, $j=1,2$, $q^{23}_m=0$, $q^{24}_m=\left(\kappa_c^2/k_d\right)K_m\left(\kappa_c a\right)$, $q^{3j}_m=-N_j\chi_jJ'_m\left(\chi_j a\right)+\left(im/a\right)M_jJ_m\left(\chi_ja\right)$, $j=1,2$, $q^{33}_m=-[im\beta/(Z_dk_da)]K_m\left(\kappa_c a\right)$, $q^{34}_m=-\left(i\kappa_c/Z_d\right)K'_m\left(\kappa_c a\right)$, $q^{4j}_m=R_jJ_m\left(\chi_ja\right)$, $j=1,2$, $q^{43}_m=-[i\kappa_c^2/(Z_dk_d)]K_m\left(\kappa_c a\right)$, and $q^{44}_m=0$. The roots $\beta$ of $\det{\mathbb Q}(\beta)=0$ yield the propagation constants of the circular GW.

\subsection{Non-circular GW via the EIE}

For the modal analysis of non-circular cores via the EIE method, we apply the latter for null external excitation \cite{Delimaris2025-pr}, i.e.,
\balg{6}
&\frac{i}{k_d}Z_d(k_d^2\mathbb{I}+\nabla\nabla^T) \oint\limits_{\xt' \in \partial V}\G_d(k_c;\xt,\xt')\un' \times \textbf{H}^d(\xt'){\rm d}\xt'\notag\\
&+\nabla\times\oint\limits_{\xt' \in \partial V}\G_d(k_c;\xt,\xt')\un' \times \textbf{E}^d(\xt'){\rm d}\xt'\!=0,\,\xt\in V.
\ealg
In \eqref{1}, $Z_d=(\mu_d/\epsilon_d)^{1/2}$ is the impedance of the cladding, $\I$ is the unity dyadic, $T$ denotes transposition, and $\G_d(k_c;\xt,\xt')$ is the tensorial form of the unbounded $V_d$ space's Green's function (GF) \cite[eq.~(A.1)]{Zouros2025-ok}, with argument $k_c=(k_d^2-\beta^2)^{1/2}=i\kappa_c$. Equation~\eqref{6} is a vectorial IR with $\xt\in V$. To proceed, we substitute in \eqref{6} the $\rho<\rho'$ branch of $\G_d(k_c;\xt,\xt')$, and apply the unprimed differential operators $k_d^2\mathbb{I}+\nabla\nabla^T$, $\nabla\times$ on $\G_d(k_c;\xt,\xt')$. This procedure yields two scalar IRs, one that is obtained from the $z$-component of \eqref{6}, and one from the $\rho-$ or the $\varphi$-component of \eqref{6} (both yield the same scalar IR). Using stacked notation, these scalar IRs read
\balg{7}
iJ_{\substack{1\\2}m}-Z_dJ_{\substack{4\\3}m}=0.
\ealg
The top part of \eqref{7}, i.e., the equation obtained using the top indices in the stacked notation, corresponds to the $z$-component of \eqref{6}, while the bottom part of \eqref{7}, i.e., the equation obtained using the bottom indices in the stacked notation, corresponds to the $\rho$- or the $\varphi$-component of \eqref{6}. The $J_{km}$, $k=1,2,3,4$, are integrals, with $J_{1m}$ and $J_{3m}$ given by
\balg{8}
J_{\substack{1\\3}m}=\oint\limits_{\xt' \in \partial V}{\mathbf M}^{(6)\ast T}_m(\kappa_c,\xt')\un'\times\begin{matrix}{\mathbf E}^d(\xt')\\[-2pt]{\mathbf H}^d(\xt')\end{matrix}{\rm d}\xt',
\ealg
where the asterisk denotes complex conjugation. The $J_{2m}$ and $J_{4m}$ are obtained, respectively, from $J_{1m}$ and $J_{3m}$, by replacing ${\mathbf M}^{(6)\ast T}_m\rightarrow{\mathbf N}^{(6)\ast T}_m$. The CVWFs appearing in \eqref{8} stem from the tensorial expansion of $\G_d(k_c;\xt,\xt')$. In particular, instead of the ${\mathbf M}^{(3)\ast T}_m$ and ${\mathbf N}^{(3)\ast T}_m$ in \cite[eq.~(A.1)]{Zouros2025-ok}, we use the ${\mathbf M}^{(6)\ast T}_m$ and ${\mathbf N}^{(6)\ast T}_m$, utilizing the properties $H^{(1)}_m(k_c\rho)=(2/\pi) i^{-m-1}K_m(\kappa_c\rho)$ and $k_cH^{(1)\prime}_m(k_c\rho)=(2/\pi) i^{-m-1}\kappa_cK'_m(\kappa_c\rho)$, where $H^{(1)}_m(x)$ is the Hankel function of the first kind and $H^{(1)\prime}_m(x)$ is its derivative with respect to the argument.

To proceed with the construction of the determinantal equation, we re-formulate \eqref{7} by suitably employing the BCs on $\partial V$. In particular, we write
\balg{9}
iJ'_{1m}-Z_dJ_{4m}&=0,\notag\\
iJ_{1m}-Z_dJ'_{4m}&=0,\notag\\
iJ'_{2m}-Z_dJ_{3m}&=0,\notag\\
iJ_{2m}-Z_dJ'_{3m}&=0,
\ealg
where we have applied the BC $\un'\times{\mathbf E}^d(\xt')=\un'\times{\mathbf E}(\xt')$, $\xt'\in\partial V$, in the primed IRs $J'_{1m}$ and $J'_{2m}$, and the BC $\un'\times{\mathbf H}^d(\xt')=\un'\times{\mathbf H}(\xt')$, $\xt'\in\partial V$, in the primed IRs $J'_{4m}$ and $J'_{3m}$. The application of the above scheme \eqref{9} is necessary because, this way, the core fields ${\mathbf E}$, ${\mathbf H}$ and the cladding fields ${\mathbf E}^d$, ${\mathbf H}^d$, get properly coupled, in a way similar to the one used in the construction of the characteristic equation for the circular isotropic DWs \cite{marcuse_2ed}.

Substituting ${\mathbf E}(\xt)$, ${\mathbf H}(\xt)$, ${\mathbf E}^d(\xt)$, and ${\mathbf H}^d(\xt)=-i/(\omega\mu_d)\nabla\times{\mathbf E}^d(\xt)$ into \eqref{9}, via \eqref{2}, \eqref{4} and \eqref{5}, we obtain the following infinite sets of linear homogeneous equations, i.e.,
\balg{10}
&\sum_{\mu=-\infty}^\infty\sum_{j=1}^2I'_{1, j \mu m}A_{j\mu}\notag\\
&+\frac{\kappa_c^2}{k_d}\sum_{\mu=-\infty}^\infty \Big(I^{(2)}_{2, \mu m}B_\mu+I^{(1)}_{2, \mu m}\Gamma_\mu\Big)=0,\notag\\
&-Z_d\sum_{\mu=-\infty}^\infty\sum_{j=1}^2I'_{4, j \mu m}A_{j\mu}-i\frac{\kappa_c^2}{k_d}\sum_{\mu=-\infty}^\infty I_{1, \mu m}\Gamma_\mu=0,\notag\\
&\sum_{\mu=-\infty}^\infty\sum_{j=1}^2I'_{2, j \mu m}A_{j\mu}-\frac{\kappa_c^2}{k_d}\sum_{\mu=-\infty}^\infty I_{1, \mu m}B_\mu=0,\notag\\
&-Z_d\sum_{\mu=-\infty}^\infty\sum_{j=1}^2I'_{3, j \mu m}A_{j\mu}\notag\\
&+i\frac{\kappa_c^2}{k_d}\sum_{\mu=-\infty}^\infty \Big(I^{(1)}_{2, \mu m}B_\mu+I^{(2)}_{2, \mu m}\Gamma_\mu\Big)=0.
\ealg
In \eqref{10}, $I'_{k, j \mu m}$, $k=1,2,3,4$, $I_{1,\mu m}$, $I^{(1)}_{2,\mu m}$, and $I^{(2)}_{2,\mu m}$, are integrals given in Appendix~\ref{APP_A}.

Equation~\eqref{10} constitutes the homogeneous system ${\mathbb A}(\beta){\mathbf u}=0$ with unknown vector ${\mathbf u}=[A_{1\mu}, A_{2\mu}, B_\mu, \Gamma_\mu]^T$ and system matrix ${\mathbb A}$ of size $4(2M+1)\times4(2M+1)$, with $M$ being the truncation bound for index $\mu$. The propagation constants $\beta$ are the roots of the determinantal equation $\det{\mathbb A}(\beta)=0$.

\subsection{Non-circular GW via the CEM}

To develop the CEM for the modal analysis of non-circular cores, we begin with the enforcement of the BCs $\un\times[{\mathbf E}^d(\xt)-{\mathbf E}(\xt)]=0$ and $\un\times[{\mathbf H}^d(\xt)-{\mathbf H}(\xt)]=0$, $\xt\in\partial V$. Expressing $\un=1/N(\varphi)[\urho-\rho'(\varphi)/\rho(\varphi)\uphi]$, $0\leqslant\varphi<2\pi$, where $N(\varphi)=\{1+[\rho'(\varphi)/\rho(\varphi)]^2\}^{1/2}$, $\rho'(\varphi)\equiv{\rm d}\rho(\varphi)/{\rm d}\varphi$, and substituting ${\mathbf E}(\xt)$, ${\mathbf H}(\xt)$, ${\mathbf E}^d(\xt)$, and ${\mathbf H}^d(\xt)=-i/(\omega\mu_d)\nabla\times{\mathbf E}^d(\xt)$ into the above BCs via \eqref{2}, \eqref{4} and \eqref{5}, we arrive at four scalar equations, i.e.,
\balg{11}
&\begin{matrix}E^d_z(\varphi)\\H^d_z(\varphi)\end{matrix}-\begin{matrix}E_z(\varphi)\\H_z(\varphi)\end{matrix}\equiv\begin{matrix}E_Z(\varphi)\\H_Z(\varphi)\end{matrix}=0,\notag\\
&\frac{1}{N(\varphi)}\Big\{\frac{\rho'(\varphi)}{\rho(\varphi)}\Big[\begin{matrix}E^d_\rho(\varphi)\\H^d_\rho(\varphi)\end{matrix}-\begin{matrix}E_\rho(\varphi)\\H_\rho(\varphi)\end{matrix}\Big]+\begin{matrix}E^d_\varphi(\varphi)\\H^d_\varphi(\varphi)\end{matrix}-\begin{matrix}E_\varphi(\varphi)\\H_\varphi(\varphi)\end{matrix}\Big\}\notag\\
&\equiv\begin{matrix}E_T(\varphi)\\H_T(\varphi)\end{matrix}=0.
\ealg
The BCs \eqref{11} are of the form $G(\varphi)=0$, with $G(\varphi)=E_Z(\varphi),~E_T(\varphi),~H_Z(\varphi),~H_T(\varphi)$. The CEM is based on the expansion of these BCs using Chebyshev polynomials. In particular, $G(\varphi)$ is approximated on the so-called Chebyshev nodes $t_k=\cos(\pi(k+1/2)/K)\in[-1,1]$, $k=0,1,\ldots,K-1$, by
\balg{12}
G(t_k)=\sum_{p=0}^{K-1}a_pT_p(t_k),\quad k=0,1,\ldots,K-1.
\ealg
In \eqref{12}, $a_p$ are expansion coefficients and $T_p(t_k)=\cos(p\pi(k+1/2)/K)$ is the Chebyshev polynomial of the first kind \cite{mason}. Utilizing the discrete orthogonality relation \cite{mason}
\balg{13}
\!\!\!\!\sum_{k=0}^{K-1}T_p(t_k)T_q(t_k)\!=\!\begin{cases}K\delta_{pq},\,p,q=0\\\frac{K}{2}\delta_{pq},\,p,q\neq0\end{cases}\!\!\!\!\!\!,K>\max\{p,q\},
\ealg
where $\delta_{pq}$ is Kronecker's delta, $a_p$ are calculated by
\balg{14}
a_p=\frac{\varepsilon_p}{K}\sum_{k=0}^{K-1}G(t_k)T_p(t_k),\quad p=0,1,\ldots,K-1,
\ealg
where $\varepsilon_p$ ($\varepsilon_0=1$, $\varepsilon_p=2$, $p\geqslant1$) is the Neumann factor. In order for \eqref{12} to be zero at every Chebyshev node $t_k$, $a_p$ should vanish for all $p=0,1,\ldots,K-1$. This yields the following linear independent equations for the construction of the homogeneous system, i.e.,
\balg{15}
\varepsilon_p\sum_{k=0}^{K-1}E_Z(t_k)T_p(t_k)&=0,\notag\\
\varepsilon_p\sum_{k=0}^{K-1}E_T(t_k)T_p(t_k)&=0,\notag\\
\varepsilon_p\sum_{k=0}^{K-1}H_Z(t_k)T_p(t_k)&=0,\notag\\
\varepsilon_p\sum_{k=0}^{K-1}H_T(t_k)T_p(t_k)&=0,
\ealg
for $p=0,1,\ldots,K-1$. Substituting, next, the expressions for $E_Z(t_k)$, $E_T(t_k)$, $H_Z(t_k)$ and $H_T(t_k)$ via \eqref{11}, and then the field components using the electric and magnetic SUPER-CVWFs, as well as the standard CVWFs, we arrive at four sets of infinite linear homogeneous equations that read
\balg{16}
&\sum_{m=-\infty}^\infty\sum_{j=1}^2 M^1_{jmp}A_{jm}+\sum_{m=-\infty}^\infty M^1_{4mp}\Gamma_{m}=0,\notag\\
&\sum_{m=-\infty}^\infty\sum_{j=1}^2 M^2_{jmp}A_{jm}\notag\\
&+\sum_{m=-\infty}^\infty \Big(M^2_{3mp}B_{m}+M^2_{4mp}\Gamma_{m}\Big)=0,\notag\\
&\sum_{m=-\infty}^\infty\sum_{j=1}^2 M^3_{jmp}A_{jm}-\frac{i}{Z_d}\sum_{m=-\infty}^\infty M^3_{3mp}B_{m}=0,\notag\\
&\sum_{m=-\infty}^\infty\sum_{j=1}^2 M^4_{jmp}A_{jm}\notag\\
&-\frac{i}{Z_d}\sum_{m=-\infty}^\infty \Big(M^4_{3mp}B_{m}+M^4_{4mp}\Gamma_{m}\Big)=0.
\ealg
In \eqref{16}, $M^{r}_{smp}$, $r,s=1,2,3,4$, are summations given in Appendix~\ref{APP_B}.

Each equation in \eqref{16} is evaluated for $p=0,1,\ldots,K-1$. In general, $K>2M+1$, where $M$ is the truncation bound for index $m$. However, in order for \eqref{16} to constitute a square homogeneous system, index $p$ should acquire the same number of values as index $m$, i.e., $2M+1$. Therefore, $p=0,1,\ldots,2M$. Concluding, \eqref{16} constitutes the homogeneous system ${\mathbb B}(\beta){\mathbf v}=0$ with unknown vector ${\mathbf v}=[A_{1m}, A_{2m}, B_m, \Gamma_m]^T$ and system matrix ${\mathbb B}$ of size $4(2M+1)\times4(2M+1)$. The propagation constants $\beta$ via the CEM are the roots of the determinantal equation $\det{\mathbb B}(\beta)=0$.

\section{Validation, Convergence, and Performance}\label{VAL}

\subsection{Validation}

\begin{figure}[!t]
\centering
\includegraphics[scale=1.0]{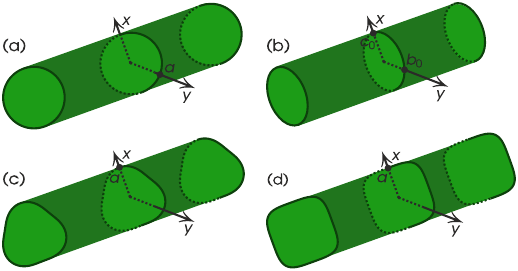}
\caption{
Circular and non-circular GWs as considered in the analysis: (a) circular; (b) elliptical; (c) rounded-triangular; (d) rounded-square.
}
\label{dw}
\end{figure}

We extensively validate the EIE and the CEM for circular and various non-circular GWs, depicted in \fig{dw}. For circular isotropic cores, the EIE, the CEM, and the SVM from Section~\ref{SOL-B} are compared against the well-known analytical solution \cite{marcuse_2ed}. Results from COMSOL are also compared with those in \cite{marcuse_2ed} for consistency. For circular gyrotropic cores, the EIE and the CEM are compared against the SVM from Section~\ref{SOL-B}, as well as with COMSOL solutions. In addition, in the special case of small non-diagonal permittivity elements of circular gyroelectric cores, the EIE and the CEM are compared with the SVM and COMSOL, but also with the approximate solution \cite{Uzunoglu1981-yq}. For non-circular cores, the EIE and the CEM are exclusively compared with COMSOL. In all examples, we compute the normalized $\beta/k_0$---the free space wavenumber $k_0$ is set equal to $1~{\rm m}^{-1}$ for simplicity---in five decimal places in order to extract valid arguments concerning the accuracy. A discussion on the performance of all methods is discussed in a subsequent Section. The parameter values for each example are provided in the table captions.

\begin{table}[!t]
\caption{$\beta/k_0$ for circular isotropic DW.\linebreak
Core: $\epsilon=1.8^2$, $\epsilon_{a}=0$, $\epsilon_{z}=1.8^2$, $\mu=1$, $\mu_{a}=0$, $\mu_{z}=1$.\linebreak
Cladding: $\epsilon_{dr}=1.5^2$, $\mu_{dr}=1$.\linebreak
$k_0a=\pi$.
}
\label{tab1}
\centering
\vspace{-4mm}
\rule{\linewidth}{1pt}

\begin{tabular*}{\linewidth}{@{ }@{\extracolsep{\fill}} r|rccc}
No & $m, p$	& EIE/CEM/SVM & Analytical~\cite{marcuse_2ed} & COMSOL  \\
\hline
1 & $\pm$2, 1	& 1.55187                 & 1.55187                & 1.55186 \\
2 & 0, 1	    & 1.55742                 & 1.55742                & 1.55740 \\
3 & 0, 2	    & 1.57022                 & 1.57022                & 1.57021 \\
4 & $\pm$1, 1	& 1.69937                 & 1.69937                & 1.69937 \\
\end{tabular*}
\rule{\linewidth}{1pt}
\end{table}

In \tab{tab1}, we consider the simple case of a circular isotropic DW. The results from the EIE, the CEM, and the SVM are compared with the well-known analytical solution \cite{marcuse_2ed}, and with COMSOL. Each propagating mode is denoted by indices $m$ and $p$, where $p$ is the order of the root. In this example, the indices $m$ and $p$ are known from \cite{marcuse_2ed}, as well as from the SVM of Section~\ref{SOL-B}. For the isotropic DW, it is known that $n_d<\beta/k_0<n$, where $n_d$ is the refractive index of the cladding and $n$ the refractive index of the core. The isotropic DW features degenerate modes, i.e., the $\beta/k_0$ have the same value for $\pm|m|$. For all four $\beta/k_0$ that exist in this configuration, all three methods, i.e., the EIE, the CEM, and the SVM, are in excellent agreement with \cite{marcuse_2ed}. COMSOL also agrees with \cite{marcuse_2ed}, and a finer mesh in COMSOL will increase its accuracy, but will also increase the execution time of its finite-element method (FEM) solver.

\begin{table}[!t]
\caption{$\beta/k_0$ for circular GEW\linebreak (small non-diagonal element).\linebreak
Core: $\epsilon=1.4$, $\epsilon_{z}=1.2$, $\mu=1$, $\mu_{a}=0$, $\mu_{z}=1$.\linebreak
Cladding: $\epsilon_{dr}=1$, $\mu_{dr}=1$.\linebreak
$k_0a=1.8$, $m=1$.}
\label{tab2}
\centering
\vspace{-4mm}
\rule{\linewidth}{1pt}

\begin{tabular*}{\linewidth}{@{ }@{\extracolsep{-2.0mm}} r|cccccc}
   &            	&     &      &     & Perturbation           &        \\
No & $\epsilon_{a}$	& EIE & CEM & SVM & \cite{Uzunoglu1981-yq} & COMSOL \\
\hline

1 & $+10^{-3}$	& 1.009479 & 1.009479 & 1.009479 & 1.009612 & 1.009478 \\
2 & $-10^{-3}$	& 1.009321 & 1.009321 & 1.009322 & 1.009489 & 1.009321 \\

3 & $+10^{-1}$	& 1.018689 & 1.018689 & 1.018689 & 1.015996 & 1.018689 \\
4 & $-10^{-1}$	& 1.003161 & 1.003161 & 1.003161 & 1.004355 & 1.003154 \\

5 & $+2\times10^{-1}$	& 1.030298 & 1.030298 & 1.030298 & 1.022791 & 1.030298 \\
6 & $-2\times10^{-1}$	& 1.000388 & 1.000388 & 1.000388 & 1.001012 & 1.000307 \\

\end{tabular*}
\rule{\linewidth}{1pt}
\end{table}

Next, in \tab{tab2}, we study the special case of a circular gyroelectric core with small non-diagonal elements in its $\boldsymbol{\epsilon}$ tensor. This case was solved by a perturbation method in \cite{Uzunoglu1981-yq}. For a GW, $n_d<\beta$ \cite{Uzunoglu1981-yq}. Here, the full-wave EIE, CEM, and SVM are compared with \cite{Uzunoglu1981-yq}, while COMSOL is also employed for consistency. In \tab{tab2}, we consider some of the small-$\epsilon_a$ cases presented in \cite[Table~1]{Uzunoglu1981-yq}, i.e., $\epsilon_a=\pm10^{-3},\pm10^{-1},\pm2\times10^{-1}$, while $m=1$. As is evident, the EIE, the CEM, and COMSOL follow the SVM. In particular, the EIE and the CEM are in agreement with the SVM up to six decimal places for all cases examined; in this example, we keep six decimal figures since the small plus/minus $\epsilon_a=10^{-3}$ values result in a change in $\beta/k_0$ evident only after the 4-th decimal digit. COMSOL is also in agreement with the SVM, however, for the $\epsilon_a=-2\times10^{-1}$ case, the agreement is up to the 4-th decimal digit. On the contrary, the perturbation method \cite{Uzunoglu1981-yq} is less accurate, as can be concluded from the presented results.

\begin{table}[!t]
\caption{$\beta/k_0$ for circular GEW\linebreak (general case).\linebreak
Core: $\epsilon=4$, $\epsilon_{a}=5.5$, $\epsilon_{z}=3$, $\mu=1$, $\mu_{a}=0$, $\mu_{z}=1$.\linebreak
Cladding: $\epsilon_{dr}=2.5$, $\mu_{dr}=1$.\linebreak
$k_0a=\pi$.
}
\label{tab3}
\centering
\vspace{-4mm}
\rule{\linewidth}{1pt}

\begin{tabular*}{\linewidth}{@{ }@{\extracolsep{\fill}} r|rcccc}
No & $m, p$ & EIE & CEM & SVM & COMSOL	\\
\hline

1 & $-2$, 1	& 1.58179 & 1.58179 & 1.58179 & 1.58145 \\
2 & $+4$, 1	& 1.69097 & 1.69097 & 1.69097 & 1.69059 \\
3 & $+1$, 1	& 2.04562 & 2.04562 & 2.04562 & 2.04515 \\
4 & $-1$, 1	& 2.18439 & 2.18439 & 2.18439 & 2.18413 \\
5 & $+3$, 1	& 2.24007 & 2.24007 & 2.24007 & 2.23990 \\

\end{tabular*}
\rule{\linewidth}{1pt}
\end{table}

In \tab{tab3}, we study a circular gyroelectric core for the general case, i.e., no restrictions apply on the $\boldsymbol{\epsilon}$ tensor, as was the case in the previous example. In this case, \cite{Uzunoglu1981-yq} cannot be applied. From the presented $\beta/k_0$ values, the EIE and the CEM are in excellent agreement with the SVM, in all five decimal places. In addition, COMSOL reproduces the SVM results with an agreement up to three decimal digits. Although an extremely fine mesh was applied in the FEM, higher accuracy is possible, but it will significantly increase the CPU time. Indices $m$ and $p$ in \tab{tab3} are calculated by the SVM. From these indices, it is understood that the circular GEW lifts the degeneracy, since the $\beta/k_0=2.04562$ that is computed for $m=+1$ has a different value from the $\beta/k_0=2.18439$ that is computed using $m=-1$.

\begin{table}[!t]
\caption{$\beta/k_0$ for circular GMW\linebreak (general case).\linebreak
Core: $\epsilon=1$, $\epsilon_{a}=0$, $\epsilon_{z}=1$, $\mu=4$, $\mu_{a}=5.5$, $\mu_{z}=3$.\linebreak
Cladding: $\epsilon_{dr}=2.5$, $\mu_{dr}=1$.\linebreak
$k_0a=\pi$.
}
\label{tab4}
\centering
\vspace{-4mm}
\rule{\linewidth}{1pt}

\begin{tabular*}{\linewidth}{@{ }@{\extracolsep{\fill}} r|rcccc}
No & $m, p$ & EIE & CEM & SVM & COMSOL	\\
\hline

1 & $+1$, 1	& 1.90586 & 1.90586 & 1.90586   & 1.90687 \\
2 & $-1$, 1	& 2.05758 & 2.05758 & 2.05758   & 2.05818 \\
3 & $+3$, 1	& 2.13972 & 2.13972 & 2.13972   & 2.14091 \\

\end{tabular*}
\rule{\linewidth}{1pt}
\end{table}

A similar example to that of \tab{tab3} is given in \tab{tab4} for a circular gyromagnetic core. Similar remarks apply here as well: the agreement between the EIE, the CEM, and the SVM is excellent, while COMSOL follows the other three methods to a satisfactory extent. The lift of degeneracy also appears here, as is evident from the computed $\beta/k_0$ which have different values for $m=\pm1$.

\begin{table}[!t]
\caption{$\beta/k_0$ for elliptical GEW and GMW.\linebreak $k_0c_0=\pi$, $k_0b_0=0.8\pi$\linebreak
Left: gyroelectric core;\linebreak
core: $\epsilon=4$, $\epsilon_{a}=5.5$, $\epsilon_{z}=3$, $\mu=1$, $\mu_{a}=0$, $\mu_{z}=1$;\linebreak
cladding: $\epsilon_{dr}=2.5$, $\mu_{dr}=1$.\linebreak
Right: gyromagnetic core;\linebreak
core: $\epsilon=1$, $\epsilon_{a}=0$, $\epsilon_{z}=1$, $\mu=4$, $\mu_{a}=0.5$, $\mu_{z}=3$;\linebreak
cladding: $\epsilon_{dr}=2$, $\mu_{dr}=1$.
}
\label{tab5}
\centering
\vspace{-4mm}
\rule{\linewidth}{1pt}

\begin{tabular*}{\linewidth}{@{ }@{\extracolsep{-2.5mm}} r|ccc|ccc}
& \;Gyroelectric          && 	& \;Gyromagnetic          && \\
No & EIE & CEM & COMSOL & EIE & CEM & COMSOL  \\
\hline

1 &  1.71274 & 1.70068   & 1.70052 &  1.47442 & 1.46538   & 1.46526 \\
2 &  1.96690 & 1.96856  & 1.96812 &  1.49180  & 1.49682 & 1.49680 \\
3 &  2.04585 & 2.09575  & 2.09556 &  1.51392  & 1.52560 & 1.52548 \\
4 & -----    & -----  & -----   &  1.67537  & 1.67867 & 1.67848 \\
5 & -----    & -----  & -----   &  1.73011  & 1.72987 & 1.72983 \\
6 & -----    & -----  & -----   &  1.75637  & 1.76510 & 1.76498 \\
7 & -----    & -----  & -----   & 1.95764   & 1.95471 & 1.95470 \\

\end{tabular*}
\rule{\linewidth}{1pt}
\end{table}

In the following, we demonstrate the validity for non-circular GWs. \tab{tab5} presents the computed $\beta/k_0$ for an elliptical cross-section core of semi-major axis $c_0$, semi-minor axis $b_0$, and eccentricity $h=[1-(b_0/c_0)^2]^{1/2}$. The left part of \tab{tab5} presents results for a gyroelectric core, while the right part presents results for a gyromagnetic core. From the presented results, the EIE is in agreement with COMSOL up to two decimal places for both the gyroelectric and gyromagnetic cases. On the contrary, the CEM matches COMSOL up to three decimal places for the gyroelectric case and up to four decimal places for the gyromagnetic case, showcasing its higher accuracy compared to the EIE.

\begin{table}[!t]
\caption{$\beta/k_0$ for rounded-triangular GEW and GMW.\linebreak $k_0a=0.8\pi$, $h=0.1$.\linebreak
Left: gyroelectric core;\linebreak
core: $\epsilon=4$, $\epsilon_{a}=0.2$, $\epsilon_{z}=3$, $\mu=1$, $\mu_{a}=0$, $\mu_{z}=1$;\linebreak
cladding: $\epsilon_{dr}=2$, $\mu_{dr}=1$.\linebreak
Right: gyromagnetic core;\linebreak
core: $\epsilon=1$, $\epsilon_{a}=0$, $\epsilon_{z}=1$, $\mu=4$, $\mu_{a}=0.2$, $\mu_{z}=3$;\linebreak
cladding: $\epsilon_{dr}=2$, $\mu_{dr}=1$.
}
\label{tab6}
\centering
\vspace{-4mm}
\rule{\linewidth}{1pt}

\begin{tabular*}{\linewidth}{@{ }@{\extracolsep{-2.5mm}} r|ccc|ccc}
& \;Gyroelectric &&& \;Gyromagnetic &&\\
No & EIE & CEM & COMSOL & EIE & CEM & COMSOL \\
\hline

1 &  1.45277 & 1.45513  & 1.45510 & 1.43335  & 1.43367 & 1.43360 \\
2 &  1.46049 & 1.46369  & 1.46376 & 1.44032  & 1.44313 & 1.44327 \\
3 &  1.50252 & 1.50689  & 1.50685 & 1.47262  & 1.47400 & 1.47387 \\
4 &  1.58370 & 1.58504  & 1.58497 & 1.62124  & 1.62214 & 1.62199 \\
5 &  1.74155 & 1.74537  & 1.74538 & 1.73232  & 1.74239 & 1.74251 \\
6 &  1.82556 & 1.83228  & 1.83227 & 1.80685  & 1.82475 & 1.82483 \\

\end{tabular*}
\rule{\linewidth}{1pt}
\end{table}

Next, in \tab{6} we consider a rounded-triangular GEW and GMW. The polar equation for the rounded-triangular cross-section is given by $\rho(\varphi)=a[h^2+2h \cos(3\varphi)+1]^{1/2}/(h+1)$, $0\leqslant\varphi<2\pi$, where $h$ is a parameter that modifies the cross-sectional shape. The EIE agrees with COMSOL up to two decimal digits for the gyroelectric case and up to three decimal digits for the gyromagnetic case. The CEM shows higher accuracy than the EIE, as compared to COMSOL, i.e., up to four decimal digits for both the gyroelectric and gyromagnetic cases.

\begin{table}[!t]
\caption{$\beta/k_0$ for rounded-square GEW and GMW.\linebreak $k_0a=0.4\pi$, $n=2$.\linebreak
Left: gyroelectric core;\linebreak
core: $\epsilon=3$, $\epsilon_{a}=0.5$, $\epsilon_{z}=5$, $\mu=1$, $\mu_{a}=0$, $\mu_{z}=1$;\linebreak
cladding: $\epsilon_{dr}=2$, $\mu_{dr}=1$.\linebreak
Right: gyromagnetic core;\linebreak
core: $\epsilon=1$, $\epsilon_{a}=0$, $\epsilon_{z}=1$, $\mu=3$, $\mu_{a}=0.5$, $\mu_{z}=5$;\linebreak
cladding: $\epsilon_{dr}=2$, $\mu_{dr}=1$.
}
\label{tab7}
\centering
\vspace{-4mm}
\rule{\linewidth}{1pt}

\begin{tabular*}{\linewidth}{@{ }@{\extracolsep{-2.5mm}} r|ccc|ccc}
& \;Gyroelectric &&& \;Gyromagnetic &&\\
No & EIE & CEM & COMSOL & EIE & CEM & COMSOL \\
\hline

1 &  1.41988 & 1.41922  & 1.41923 & 1.41911 & 1.41841 & 1.41843 \\
2 &  1.53758 & 1.53757 & 1.53750 & 1.53220 & 1.53566 & 1.53569 \\

\end{tabular*}
\rule{\linewidth}{1pt}
\end{table}

As a final demonstrative example, in \tab{tab7} we examine a rounded-square GEW and GMW. The polar equation for the rounded-square is $\rho(\varphi)=a/(\cos^{2n}\varphi+\sin^{2n}\varphi)^{1/(2n)}$, $0\leqslant\varphi<2\pi$, where $n$ is a parameter that modifies the cross-sectional shape. Regarding the parameter values used in this example, this setup allows two values of $\beta/k_0$ for both gyroelectric and gyromagnetic cores. Still, the CEM consistently demonstrates higher accuracy than the EIE, achieving four decimal digits compared to COMSOL, whereas the EIE achieves four and two decimal digits for the gyroelectric and gyromagnetic cases, respectively.

\subsection{Convergence}

\begin{figure}[!t]
\centering
\includegraphics[scale=1.0]{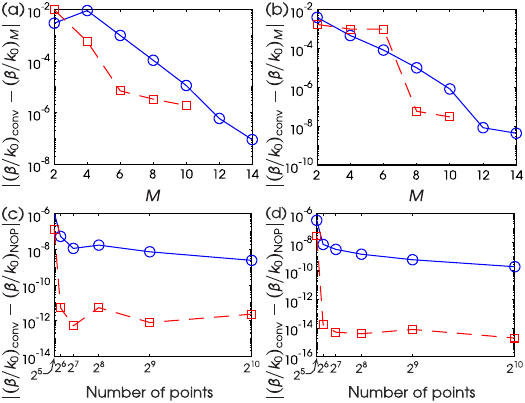}
\caption{
Convergence plots. (a)--(b) $|(\beta/k_0)_{\rm conv}-(\beta/k_0)_{M}|$ against $M$ by keeping $32$ integration points fixed for the EIE and $K=128$ for the CEM; (c)--(d) $|(\beta/k_0)_{\rm conv}-(\beta/k_0)_\text{NOP}|$ against number of points (integration points and Chebyshev nodes) by keeping fixed $M=10$ for both the EIE and CEM. Red curves/squares: EIE; blue curves/circles: CEM. (a), (c) Convergence of $(\beta/k_0)_\text{EIE}=1.71274$ and $(\beta/k_0)_\text{CEM}=1.70068$ (gyroelectric case/result No~1 in \tab{tab5}); (b), (d) Convergence of $(\beta/k_0)_\text{EIE}=1.47442$ and $(\beta/k_0)_\text{CEM}=1.46538$ (gyromagnetic case/result No~1 in \tab{tab5}).
}
\label{cn}
\end{figure}

In \fig{cn}, we examine the convergence of the EIE and CEM. We study how the values of $\beta/k_0$ converge, by computing $|(\beta/k_0)_{\rm conv}-(\beta/k_0)_{M}|$ and $|(\beta/k_0)_{\rm conv}-(\beta/k_0)_{\rm NOP}|$, where $(\beta/k_0)_{\rm conv}$ is the converged value. For the first calculation (see Figs.~\ref{cn}(a) and (b)), we keep the number of points that sample the contour $\partial V$ fixed (i.e., the order of integration for the EIE and the number $K$ of Chebyshev nodes for the CEM), and vary the value of truncation bound $M$ to obtain the $(\beta/k_0)_{M}$. For the second calculation (see Figs.~\ref{cn}(c) and (d)), we keep $M$ fixed and vary the number of points that sample the contour $\partial V$ to obtain the $(\beta/k_0)_{\rm NOP}$. Obviously, we cannot implement the above study for all $\beta/k_0$ given in \tabs{tab1}--\tab{tab7}, so to demonstrate the convergence, we perform this study for the result No~1 of \tab{tab5} (i.e., $(\beta/k_0)_\text{EIE}=1.71274$, $(\beta/k_0)_\text{CEM}=1.70068$ for the gyroelectrtic case and $(\beta/k_0)_\text{EIE}=1.47442$, $(\beta/k_0)_\text{CEM}=1.46538$ for the gyromagnetic case). Figures~\ref{cn}(a) and (c) depict the convergence for the gyroelectic case, and Figs.~\ref{cn}(b) and (d) for the gyromagnetic case.

From Figs.~\ref{cn}(a) and (b), the CEM exhibits a convergence that reaches seven and eight significant digits for the gyroelectric and gyromagnetic case, respectively, using $M=14$. The convergence of the CEM is further improved to more significant figures when higher values of $M$ are used. On the contrary, the EIE stagnates for $M\geqslant11$ and cannot achieve convergence beyond five or seven significant digits in the gyroelectric and gyromagnetic cases. For this reason, values for $M>10$ are not depicted for the EIE.

Finally, Figs.~\ref{cn}(c) and (d) show that, for a fixed $M$, the EIE can achieve convergence to twelve and fourteen significant digits for the gyroelectric and gyromagnetic case, respectively, when a high integration order is used. The CEM, on the other hand, delivers eight and nine significant figures for the gyroelectric and gyromagnetic cases, respectively. Nevertheless, the discussion in \fig{cn} concerns the self-convergence of the two methods, and the actual accuracy discussed in \tab{tab5} leads to the conclusion that the CEM achieves a better agreement with COMSOL, compared to the EIE. As a final comment on the impact of $M$ and the number of points on convergence, an increment to the number of points yields significantly improved convergence, but $M$ has a greater impact on the accuracy, especially for non-circular cores (i.e., better accuracy is obtained by keeping the number of points to a moderate value and increasing $M$, rather than keeping $M$ to a moderate value and increasing the number of points).

\begin{table}[!t]
\caption{Computational performance.}
\label{tab8}
\centering
\vspace{-4mm}
\rule{\linewidth}{1pt}

\begin{tabular*}{\linewidth}{@{ }@{\extracolsep{-1.9mm}} c|cc|cc|cc}
& \multicolumn{2}{c|}{EIE} & \multicolumn{2}{c|}{CEM} & \multicolumn{2}{c}{COMSOL} \\
        & CPU time & Memory & CPU time & Memory & CPU time & Memory \\
Example & (s) & (GB) & (s) & (GB) & (s) & (GB) \\
\hline

\tab{tab5} & 119 & 0.28 & 70 & 0.33 & 332 & 22.5 \\
\tab{tab6} & $\phantom{1}$85 & 0.28 & 63 & 0.33 & 333 & 22.9 \\
\tab{tab7} & $\phantom{1}$76 & 0.28 & 48 & 0.33 & $\phantom{3}$49 & $\phantom{2}$4.8 \\

\end{tabular*}
\rule{\linewidth}{1pt}
\end{table}

\subsection{Performance}

The computational performance, i.e., the CPU time and memory consumption of the EIE, the CEM, and COMSOL, is discussed in \tab{tab8} for the non-circular core examples of \tabs{tab5}--\ref{tab7}. To obtain the results collected in \tab{tab8}, one value of $\beta/k_0$ is computed under the constraint of five decimal digits of convergence. For this purpose, the EIE was initialized using the truncation bound $M=10$ and $128$ integration points; the CEM was initialized using $M=12,14,10$ for the examples of \tabs{tab5},\ref{tab6},\ref{tab7}, respectively, and $K=128$; finally, COMSOL was initialized using a dense mesh of $0.2$ maximum element size. In particular, we have calculated the first $\beta/k_0$ (results No~1 in \tabs{tab5}--\ref{tab7} for the gyroelectric case); similar data are obtained for the rest of the $\beta/k_0$ values, as well as for the gyromagnetic cases. To achieve five convergent decimal digits, CEM outperforms EIE and COMSOL in CPU time, and COMSOL in memory consumption; the RAM requirements for EIE and CEM are practically similar. The shape of the core is also crucial. It is indicative that COMSOL's FEM solver requires almost $23$~GB RAM for an elliptical and rounded-triangular core, compared to the reduced $0.28$~GB/$0.33$~GB of the EIE and the CEM. However, COMSOL has a reduced consumption of $4.8$~GB RAM and a similar CPU time with CEM, for the rounded-square core.

Based on the above discussion, including validation, convergence, and performance, we conclude that the CEM performs better in computing $\beta/k_0$ than the EIE and is therefore recommended as a preferable alternative for the calculation of propagation characteristics in non-circular GWs.

\section{Microwave Application: Complex Propagation Constants of Ferrite GMWs}\label{APP}

Herein, we present a microwave application of ferrite GMWs in the presence of an external magnetic flux density bias. Since a ferrite features dispersion properties with losses, the propagation constants $\beta$ are complex. We calculate the complex, normalized $\beta/k_0$ for circular and elliptical cores, validate the results with COMSOL, and discuss how introducing the eccentricity, that is equivalent to a perturbation of the core geometry, alters the $\beta/k_0$ and the modal patterns.

When a ferrite is magnetized by an external magnetic flux density ${\mathbf B}_0=B_0\uz$, it exhibits a gyromagnetic response \cite{pozar_4ed}. The $\m$ relative elements are $\mu=1+\omega_0 \omega_m/(\omega_0^2-\omega^2)$, $\mu_{a}=\omega\omega_m/(\omega_0^2-\omega^2)$ and $\mu_{z}=1$. In the present study, we consider the G-610 aluminum garnet as the ferrite material, where $\omega_m=\mu_0\gamma M_s$, $\gamma=1.759\times 10^{11} \mathrm{C/kg}$ is the gyromagnetic ratio, $4\pi M_s=680$~G \cite{pozar_4ed}, $\omega_0=\mu_0\gamma H_0-i \mu_0 \gamma \Delta H/2$ is the complex Larmor circular frequency, $H_0=B_0/\mu_0$ is the external magnetic field intensity, and $\Delta H=40~\mathrm {Oe}$ \cite{pozar_4ed}. The $\e$ is isotropic with relative values $\epsilon=\epsilon_{z}=14.5+i 0.0029$ and $\epsilon_{a}=0$ \cite{pozar_4ed}.

\begin{table}[!t]
\caption{
Complex $\beta/k_0$ values for circular G-610 aluminum garnet ferrite GMW.\linebreak
Core: G-610.\linebreak
Cladding: $\epsilon_{dr}=1$, $\mu_{dr}=1$.\linebreak
$f=4~{\rm GHz}$, $a=1~{\rm cm}$, $B_0=1~{\rm T}$.
}
\label{tab9}
\centering
\vspace{-4mm}
\rule{\linewidth}{1pt}
\begin{tabular*}{\linewidth}{@{\extracolsep{-2.8mm}}r|rllll}
No & $m, p$ & EIE & CEM & SVM & COMSOL  \cr
\hline
1  & \;$+1$, 1 & $2.97446$      & $2.97446$      & $2.97447$      & $2.97442$      \cr
   &           & $+i0.00085883$ & $+i0.00085883$ & $+i0.00085884$ & $+i0.00085883$ \cr
2  & \;$-1$, 1 & $2.93819$      & $2.93819$      & $2.93818$      & $2.93815$      \cr
   &           & $+i0.00071097$ & $+i0.00071097$ & $+i0.00071097$ & $+i0.00071096$ \cr
3  & \;$0$,  1 & $1.98134$      & $1.98134$      & $1.98134$      & $1.98129$      \cr
   &           & $+i0.00062067$ & $+i0.00062067$ & $+i0.00062068$ & $+i0.00062066$ \cr

\end{tabular*}
\rule{\linewidth}{1pt}
\end{table}

In \tab{tab9}, we present the computed values of the complex $\beta/k_0$ for a circular G-610 GMW, at an operating frequency $f=4~{\rm GHz}$, when $B_0=1~{\rm T}$. We focus on these complex $\beta/k_0$ that correspond to localized modes inside the ferrite core. Since the core is circular, we employ all methods, i.e., the EIE, the CEM, the SVM, as well as COMSOL. For the given $f$, three complex $\beta/k_0$ exist, given in \tab{tab9}. The mode indices $m$ and $p$ are calculated by the SVM. The complex $\beta/k_0$ for results No~1 and No~2 correspond to non-degenerate modes, since they are obtained for opposite $m$ values. All four methods agree up to four decimal digits for both the real and the imaginary part; for result No~2, the EIE, the CEM, and the SVM agree up to five decimal digits for the imaginary part. These remarks demonstrate the capacity of the developed methods to address lossy and dispersive GMWs.

\begin{figure}[!t]
\centering
\includegraphics[scale=1.0]{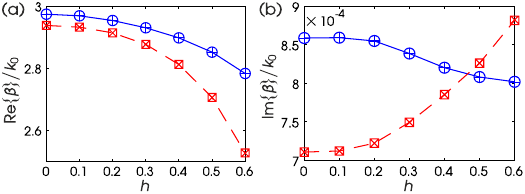}
\caption{
Complex $\beta/k_0$ against $h$ for elliptical G-610 aluminum garnet ferrite GMW. The values of the parameters are the same as those in \tab{tab9}, except for $c_0$, which is equal to $1~{\rm cm}$. (a) ${\rm Re}\{\beta\}/k_0$; (b) ${\rm Im}\{\beta\}/k_0$. Blue curves: mode~1; red curves: mode~2. Blue circles: CEM; blue plus signs: COMSOL; red squares: CEM; red crosses: COMSOL.
}
\label{cb}
\end{figure}

Our study is further extended in \fig{cb}, where we examine how the complex $\beta/k_0$ vary by introducing an elliptical core and gradually increasing its eccentricity $h$. In particular, the elliptical core has a semi-major axis $c_0=1~{\rm cm}$. Figure~\ref{cb}(a) depicts the change of ${\rm Re}\{\beta\}/k_0$ against $h$ and \fig{cb}(b) the change of ${\rm Im}\{\beta\}/k_0$ against $h$, where ${\rm Re}$ and ${\rm Im}$ denote the real and the imaginary part, respectively. We begin from a circular core ($h=0$) and reach the value $h=0.6$. Values of $h$ higher than $0.6$ yield modes whose patterns are not localized inside the ferrite core but instead extend spatially into the cladding. Specifically, the curves in \fig{cb} depict how the two non-degenerate modes of the circular G-610 GMW (results No~1 and 2 in \tab{tab9}) evolve with the change of $h$. When $h=0$, the starting values of ${\rm Re}\{\beta\}/k_0$ and ${\rm Im}\{\beta\}/k_0$ equal those given in \tab{tab9}. The blue curves in \fig{cb} correspond to the mode having $\beta/k_0=2.97446+i0.00085883$ when $h=0$ (this mode is referred hereafter as mode~1), and the red curves to the mode with $\beta/k_0=2.93819+i0.00071097$ when $h=0$ (this mode is referred hereafter as mode~2). As $h$ increases, the real part decreases for both modes. On the contrary, the imaginary part for mode~1 slightly decreases while the imaginary part for mode~2 increases. This behavior is verified by both the CEM (for the results in \fig{cb} we used the CEM solely) and COMSOL, as the results from both methods, for both the real and the imaginary part, coincide.

\begin{figure}[!t]
\centering
\includegraphics[scale=1.0]{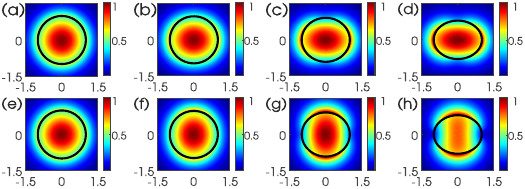}
\caption{
Modal patterns of the normalized total electric field magnitude, for various values of $h$, for the localized core-modes studied in \fig{cb}.
(a)--(d): Localized modal patterns corresponding to the complex $\beta/k_0$ for mode~1 (blue curves in \fig{cb}).
(e)--(h): Localized modal patterns corresponding to the complex $\beta/k_0$ for mode~2 (red curves in \fig{cb}).
(a) and (e) $h=0$; (b) and (f) $h=0.2$; (c) and (g) $h=0.4$; (d) and (h) $h=0.6$.
}
\label{fp}
\end{figure}

Finally, we depict in \fig{fp} the respective modal patterns of modes~1 and 2 by plotting the magnitude of the transverse total electric field ${\mathbf F}(\xt)$, where ${\mathbf F}={\mathbf E}$, $\xt\in V$ and ${\mathbf F}={\mathbf E}^d$, $\xt\in V_d$, normalized to its maximum value, for $h=0,0.2,0.4,0.6$. For $h=0,0.2$, the two modes are localized inside the G-610 core while the decaying part extends spatially inside the cladding. When $h=0.4,0.6$, the patterns are elongated; mode~1 is elongated along the major axis and mode~2 along the minor axis. In addition, these two modes enter the cladding, unlike the $h=0,0.2$ cases, where the modes are well localized. It should also be mentioned that mode~2 for $h=0.4$ does not have a strong localization, compared to mode~1 for the same value of $h$. Higher values of $h$ than those given in \fig{fp} yield patterns that are not localized and extend significantly into the cladding.

\section{Conclusion}\label{CON}

We analyzed non-circular GWs and calculated their propagation constants. To this end, we constructed two full-wave independent methods, namely, the EIE and the CEM, while an SVM was also constructed for the particular case of circular GWs. All three methods were verified by considering various circular and non-circular GWs. Their validity was established with comparisons against COMSOL Multiphysics, as well as with a perturbation-based method for circular GEWs, when the non-diagonal permittivity element is small. The convergence and the computational performance of the EIE and the CEM were also discussed, and we confirmed their capacity for revealing the propagation characteristics of non-circular GWs. Finally, the developed CEM was utilized for a microwave application regarding the calculation of complex propagation constants in circular and elliptical ferrite GMWs, under external magnetic flux
density bias.
\appendices

\setcounter{equation}{0}
\renewcommand{\theequation}{A.\arabic{equation}}
\section{EIE Integrals}\label{APP_A}

Eq.~\eqref{10} involves the integrals $I'_{k, j \mu m}$, $k=1,2,3,4$. $I'_{1, j \mu m}$ and $I'_{2, j \mu m}$ are given by
\balgnl{}
&\!\!\!I'_{1, j \mu m}=\int_0^{2\pi}\frac{e^{i(\mu-m)\varphi}}{\sqrt{1+\left(\rho'/\rho\right)^2}}W_j J_{\mu}(\chi_j \rho)\notag\\
&\!\!\!\times\left[\frac{im\rho'}{\rho^2}K_m(\kappa_c\rho)+\kappa_cK'_m(\kappa_c\rho)\right]\rho{\rm d}\varphi,\notag\\
&\!\!\!I'_{2, j \mu m}=\int_0^{2\pi}\frac{e^{i(\mu-m)\varphi}}{\sqrt{1+\left(\rho'/\rho\right)^2}}W_j J_{\mu}(\chi_j \rho)\notag\\
&\!\!\!\times\left[\frac{i\beta\rho'\kappa_c}{k_d\rho}K'_m(\kappa_c\rho)+\frac{m\beta}{k_d\rho}K_m(\kappa_c\rho)\right]\rho{\rm d}\varphi%\notag\\
\ealgnl
\balg{A1}
&\!\!\!-\frac{\kappa^2_c}{k_d}\int_0^{2\pi}\!\!\!\frac{e^{i(\mu-m)\varphi}}{\sqrt{1+\left(\rho'/\rho\right)^2}}\Big\{\!\!-T_j\chi_jJ'_{\mu}(\chi_j\rho)+\frac{imS_j}{\rho}J_{\mu}(\chi_j\rho)\notag\\
&\!\!\!+\frac{\rho'}{\rho}\Big[S_j\chi_jJ'_{\mu}(\chi_j\rho)+\frac{imT_j}{\rho}J_{\mu}(\chi_j\rho)\Big]\Big\}K_m(\kappa_c\rho)\rho{\rm d}\varphi,
\ealg
while $I'_{3, j \mu m}$ and $I'_{4, j \mu m}$ are given by $I'_{1, j \mu m}$ and $I'_{2, j \mu m}$, respectively, by replacing $W_j\rightarrow R_j$, $T_j\rightarrow N$ and $S_j\rightarrow M_j$.

Eq.~\eqref{10} also involves the integrals $I_{1,\mu m}$, $I^{(1)}_{2,\mu m}$, and $I^{(2)}_{2,\mu m}$, given by
\balg{A2}
&I_{1, \mu m} = \int_0^{2\pi}\frac{e^{i(\mu-m)\varphi}}{\sqrt{1+\left(\rho'/\rho\right)^2}}K_{\mu}\left(\kappa_c \rho\right)\notag\\
&\times\left[\frac{i m \rho'}{\rho^2}K_m\left(\kappa_c \rho\right) +\kappa_c K'_m\left(\kappa_c \rho\right)\right]\rho{\rm d}\varphi,\notag\\
&I^{(1)}_{2, \mu m}= \int_0^{2\pi}\frac{e^{i(\mu-m)\varphi}}{\sqrt{1+\left(\rho'/\rho\right)^2}} K_m\left(\kappa_c \rho \right)\notag\\
&\times\left[\kappa_c K'_{\mu}\left(\kappa_c \rho\right)-\frac{i\mu\rho'}{\rho^2}K_{\mu}\left(\kappa_c \rho\right)\right]\rho{\rm d}\varphi,\notag\\
&I^{(2)}_{2, \mu m}= -\int_0^{2\pi}\frac{e^{i(\mu-m)\varphi}}{\sqrt{1+\left(\rho'/\rho\right)^2}} K_{\mu}\left(\kappa_c \rho \right)\notag\\
&\times \left[\frac{i \beta \kappa_c \rho'}{k_d\rho}K'_m\left(\kappa_c \rho\right)+\frac{m\beta}{k_d\rho}K_m\left(\kappa_c \rho\right)\right]\rho{\rm d}\varphi \notag\\
&+\int_0^{2\pi}\frac{e^{i(\mu-m)\varphi}}{\sqrt{1+\left(\rho'/\rho\right)^2}}K_m\left(\kappa_c \rho\right)\notag\\
&\times\left[\frac{\mu\beta}{k_d\rho}K_{\mu}\left(\kappa_c\rho\right)-\frac{i \beta\kappa_c\rho'}{k_d\rho}K'_{\mu}\left(\kappa_c\rho\right)\right]\rho{\rm d}\varphi.
\ealg

\setcounter{equation}{0}
\renewcommand{\theequation}{B.\arabic{equation}}
\section{CEM Summations}\label{APP_B}

Eq.~\eqref{16} involves the summations $M^{r}_{smp}$, $r,s=1,2,3,4$. The summations $M^1_{jmp}$, $M^2_{jmp}$, $M^1_{4mp}$, $M^2_{3mp}$, and $M^2_{4mp}$ are given by
\balgnl
&M^1_{jmp} = -W_j\varepsilon_p\sum_{k=0}^{K-1} T_p(t_k)J_m(\chi_j\rho_k)e^{i m \varphi_k},\notag\\
&M^2_{jmp} = -\varepsilon_p\sum_{k=0}^{K-1}T_p(t_k)\Big\{\frac{1}{N(\varphi_k)}\frac{\rho'_k}{\rho_k} \Big[T_j\frac{im}{\rho_k}J_m(\chi_j\rho_k)\notag\\
&+\chi_j S_j J'_m(\chi_j\rho_k)\Big] -T_j\chi_jJ'_m(\chi_j\rho_k)\notag\\
&+\frac{imS_j}{\rho_k}J_m(\chi_j\rho_k)\Big\}e^{im\varphi_k},\notag\\
&M^1_{4mp} = -\frac{\kappa_c^2}{k_d}\varepsilon_p\sum_{k=0}^{K-1} T_p(t_k)K_m(\kappa_c\rho_k)e^{i m \varphi_k},\notag\\
&M^2_{3mp} = \varepsilon_p\sum_{k=0}^{K-1}T_p(t_k)\Big[\frac{1}{N(\varphi_k)}\frac{\rho'_k}{\rho_k} \frac{im}{\rho_k}K_m(\kappa_c\rho_k)\notag\\
&-\kappa_c K'_m(\kappa_c\rho_k)\Big]e^{im\varphi_k},%\notag\\
\ealgnl
\balg{B1}
&M^2_{4mp} = \varepsilon_p\sum_{k=0}^{K-1}T_p(t_k)\Big[\frac{1}{N(\varphi_k)}\frac{\rho'_k}{\rho_k} \frac{i\beta}{k_d}K'_m(\kappa_c\rho_k)\notag\\
&-\frac{m\beta}{k_d\rho_k}K_m(\kappa_c\rho_k)\Big]e^{im\varphi_k},
\ealg
where $\varphi_k=\pi(1+t_k)$. $M^3_{jmp}$ is given by $M^1_{jmp}$ by replacing $W_j\rightarrow R_j$, the $M^4_{jmp}$ is given by $M^2_{jmp}$ by replacing $T_j\rightarrow N_j$ and $S_j\rightarrow M_j$, $M^3_{3mp}=M^1_{4mp}$, $M^4_{3mp}=M^2_{4mp}$, and $M^4_{4mp}=M^2_{3mp}$.

{\small

}

\end{document}